\documentclass[10pt]{article}
\usepackage{geometry}
\usepackage[utf8]{inputenc}
\usepackage[version=3]{mhchem}

\usepackage{hyperref}
\hypersetup{
    colorlinks,
    citecolor=black,
    filecolor=black,
    linkcolor=black,
    urlcolor=black
}

\usepackage{titlesec}
\usepackage{stmaryrd}
\usepackage{float}
\usepackage{soul}
\usepackage{amsmath}
\usepackage{multicol}

\usepackage{subcaption}
\usepackage[export]{adjustbox}
\usepackage{wrapfig}
\usepackage{graphicx}

\usepackage[backend=biber,style=chem-acs,maxnames=99,giveninits=true, sorting=none]{biblatex}
\DeclareDelimFormat[bib,biblist]{multicitedelim}{\addcomma} 
\DeclareDelimFormat{multicitedelim}{\addcomma} 
\renewcommand{\cite}[1]{\textsuperscript{\autocite{#1}}}

\DeclareBibliographyDriver{article}{%
  \usebibmacro{author}%
  \setunit{\labelnamepunct}\newblock
  \usebibmacro{title}%
  \newblock
  \printfield{journaltitle}%
  \setunit{\space}%
  \textbf{\printfield{year}}%
  \setunit*{\addcomma\space}%
  \printfield{volume}%
  \setunit*{\addcomma\space}%
  \printfield{pages}%
  \setunit{\finentry\\}%
  \printfield{doi}%
  \finentry
}
\DeclareBibliographyDriver{misc}{%
  \usebibmacro{author} 
  \setunit{\labelnamepunct}\newblock
  \usebibmacro{title}%
  \newblock
  \textit{, Preprint} 
  \setunit{\addspace}%
  \textbf{\printfield{year}} 
  \setunit{\finentry\\}%
  \printfield{url}%
  \finentry
}
\DeclareBibliographyDriver{inproceedings}{%
  \usebibmacro{author}%
  \setunit{\labelnamepunct}\newblock
  \textnormal{\usebibmacro{title}}%
  \newblock
  \printfield{booktitle}%
  \setunit{\space}%
  \textbf{\printfield{year}}%
  \setunit*{\addcomma\space}%
  \printfield{volume}%
  \setunit*{\addcomma\space}%
  \printfield{pages}%
  \finentry
}
\DeclareBibliographyDriver{software}{%
  \printfield{url}%
  \finentry
}

\graphicspath{{./figures}}

\title{Systematic incorporation of nuclear quantum effects into atomistic simulations by smoothed trajectory analysis}

\begin{document}
\date{}
\author{}
\maketitle
\vspace{-5em}
\begin{center}
\large
Ádám Madarász$^{1,}$*, Bence Balázs Mészáros$^{2,3}$, János Daru$^{3,}$*\\
\vspace{1em}
\small\textit{
Research Centre for Natural Sciences, Magyar Tudóosok Körútja 2, H-1117 Budapest, Hungary\\
Hevesy György PhD School of Chemistry Eötvös Loránd University, Pázmány Péter sétány 1/A, 1117 Budapest, Hungary\\
Department of Organic Chemistry, Eötvös Loránd University, Pázmány Péter sétány 1/A, 1117 Budapest, Hungary\\}
\vspace{1em}
E-mail: madarasz.adam@ttk.hu, janos.daru@ttk.elte.hu
\end{center}

\section{Abstract}

Nuclear quantum effects (NQEs) play an essential role in many atomistic systems, yet their explicit inclusion in molecular simulations remains challenging. Path-integral molecular dynamics (PIMD) provides a rigorous framework for incorporating NQEs, but its practical applicability is often limited by the slow and strongly system-dependent convergence with respect to the number of beads.
Here we introduce path-integral generalized smoothed trajectory analysis (PIGSTA), a post-processing framework for the systematic incorporation of NQEs into atomistic simulations, using either classical or path-integral molecular dynamics trajectories. By applying analytically defined convolution kernels to simulation trajectories, PIGSTA corrects the frequency-dependent discretization error associated with a finite number of beads, without modifying the underlying dynamics.
For harmonic systems, PIGSTA recovers the exact quantum-mechanical limit at any bead number, whereas standard PIMD becomes exact only in the infinite-bead limit. More generally, the method significantly improves the convergence of thermodynamic and structural observables at finite bead numbers and provides an internal, reference-free diagnostic of bead-number convergence based on the consistency of energy and force estimators.
We assess PIGSTA for ambient liquid water and for the Zundel cation at ultralow temperature, representing a particularly demanding case for bead-number convergence. In both systems, PIGSTA reproduces the converged PIMD limit and enables physically consistent results at reduced bead numbers when convergence criteria are satisfied. Owing to its post-processing nature and negligible additional computational cost, PIGSTA offers a practical and broadly applicable approach for incorporating NQEs into atomistic simulations.
\\
\hrule

\vspace{6em}
\begin{multicols}{2}
\section{Introduction}

Quantum fluctuations of the nuclei give rise to a wide range of physical and chemical
phenomena across broad temperature regimes, including the structural and thermophysical
properties of water,\cite{Ceriotti2016-no,Markland2018-nv} various organic liquids,\cite{Ugur2025-ym} quantum tunneling,\cite{Cahlik2021-ix,Heller2021-uv,Olajide2024-on} proton transport in zeolites,\cite{Bocus2023-ng} and vibrational and spectroscopic properties.\cite{Engel2021-du,Tsuru2025-yl} In large and complex condensed-phase systems and materials, explicit quantum-mechanical treatments of the nuclear degrees of freedom rapidly become computationally infeasible,\cite{Pavosevic2020-ku} leaving path-integral–based approaches as the only generally applicable framework for incorporating NQEs at finite temperature. PIMD provides a formally exact description of quantum nuclear fluctuations in the limit of an infinite number of beads,\cite{Feynman2005-mq,Berne1986-jf} with a computational cost that scales linearly with the bead number on top of the underlying interaction model, making it particularly attractive for simulations at the length and time scales relevant to computational materials science.

Despite this favorable formal scaling, the practical applicability of PIMD is often limited by the slow and strongly system- and property-dependent convergence with respect to the number of beads. In realistic simulations, different observables may converge at markedly different rates, making it nontrivial to determine when a simulation has reached convergence in a physically meaningful sense. Crucially, there is no straightforward, system-independent criterion to assess whether a given PIMD simulation is sufficiently converged, nor a general way to diagnose insufficient convergence without resorting to substantially increasing the bead number or performing costly reference calculations.

To alleviate bead-number convergence challenges in path-integral molecular dynamics, several acceleration schemes have been developed, most notably the path-integral
generalized Langevin equation thermostat (PIGLET)\cite{Ceriotti2012-ei} and the
path-integral quantum thermal bath (PIQTB).\cite{Brieuc2016-wy} These approaches exploit generalized Langevin dynamics,\cite{Ceriotti2009-gl,Ceriotti2011-ka} colored-noise thermostats,\cite{Ceriotti2009-yq,Ceriotti2010-cd} or quantum thermal baths\cite{Dammak2009-zz} to reproduce the correct quantum fluctuations of
harmonic systems using a reduced number of beads. While highly effective in many applications, their performance relies on assumptions that are formally exact in the harmonic limit and whose validity can be more difficult to assess in strongly anharmonic or extreme conditions,\cite{Schran2018-py,Uhl2016-jb} where bead-number convergence can become non-monotonic and strongly property-dependent.

GSTA offers an alternative route to incorporate quantum fluctuations by filtering classical trajectories in a post-processing step, without modifying the underlying dynamics.\cite{Berta2020-gz} In this work, we extend this concept to path-integral molecular dynamics and introduce PIGSTA. By applying analytically defined convolution kernels derived from the quantum harmonic-oscillator model and explicitly dependent on the bead number, PIGSTA retains exactness in the harmonic limit while providing an alternative strategy to improve and assess bead-number convergence in path-integral simulations. As we show below, this formulation naturally enables an internally consistent, reference-free assessment of convergence. The conceptual idea underlying the PIGSTA framework is illustrated schematically in Figure~\ref{fgr:concept}.

\newcommand{\figwidth}[0]{0.95\linewidth}
\begin{figure}[H]
\centering
\includegraphics[width=\figwidth]{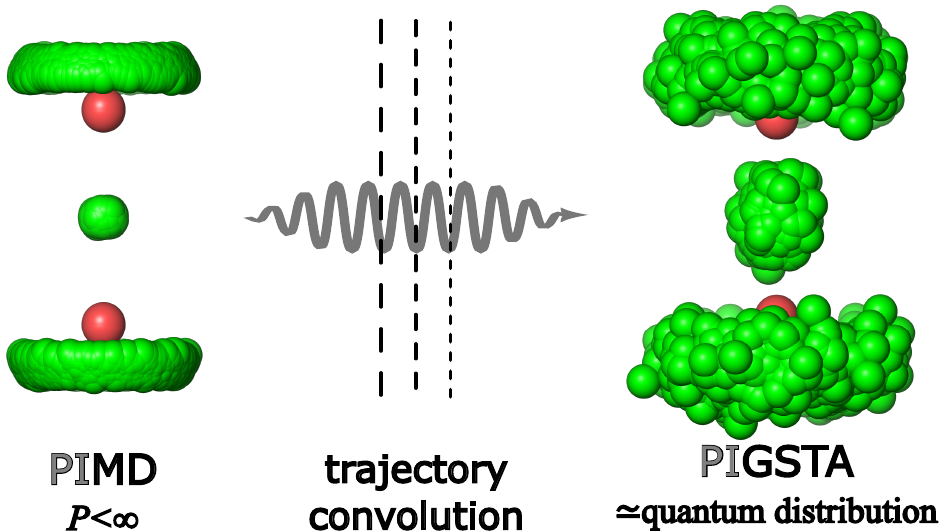}
\caption{\textbf{Schematic overview of the trajectory convolution framework underlying GSTA and its path-integral extension.}
The same trajectory-based convolution framework applies to both classical molecular dynamics (MD/GSTA) and path-integral molecular dynamics (PIMD/PIGSTA). Left: representative nuclear configurations from a (PI)MD simulation of the Zundel cation (H$_5$O$_2^+$) at finite bead number. Middle: schematic representation of the frequency-dependent trajectory convolution operator. Right: the resulting PIGSTA-filtered configurations, yielding a statistical representation closer to the converged quantum distribution without modifying the underlying dynamics. All configurations are aligned by fitting the oxygen atoms prior to visualization.}
\label{fgr:concept}
\end{figure}

The behaviour of PIGSTA is most clearly revealed in regimes where bead-number convergence poses a particular challenge. Ultralow-temperature, strongly anharmonic hydrogen-bonded systems, such as the Zundel cation, represent extreme cases in which very large bead numbers may be required to obtain stable structural descriptors using standard approaches. We therefore adopt the Zundel cation at ultralow temperature as a stringent stress test to benchmark PIGSTA against standard PIMD, PIGLET, and PIQTB. To demonstrate that the proposed approach is not limited to such extreme molecular cluster ions, we further assess PIGSTA for ambient liquid water, a widely used benchmark system in path-integral simulations.

PIGSTA is a post-processing framework that filters path-integral trajectories using analytically defined convolution kernels to improve and assess bead-number convergence in PIMD simulations. It provides a parameter-free and code-independent alternative to established acceleration schemes such as PIGLET and PIQTB, while retaining exactness in the harmonic limit and converging to the exact quantum result in the infinite-bead limit.

Beyond improving numerical convergence, the post-processing nature of PIGSTA enables an internal, reference-free assessment of physical consistency based on the agreement between formally distinct energy and force estimators. This dual role—combining improved convergence with built-in diagnostics of harmonicity and internal consistency—distinguishes PIGSTA from thermostat-based approaches and allows the identification of bead-number regimes in which path-integral results can be considered physically meaningful.

\section{Theory}

In the following, we briefly recall the path-integral formulation of quantum statistical mechanics and introduce the mathematical framework of the proposed PIGSTA method, which extends GSTA to PIMD.

\subsection{Path-integral formulation of nuclear quantum statistics}

In the path-integral formulation of quantum statistical mechanics, the partition function of a system of nuclei can be expressed as the classical partition function of a ring polymer consisting of $P$ replicas of the physical system reads
\begin{equation}
\mathcal{H}_P(\mathbf{v},\mathbf{x}) = \sum_{k=0}^{P-1} \left(
\frac{1}{2} m v_k^2
+ V(x_k)
+ \frac{1}{2} m \omega_P^2 (x_k - x_{k+1})^2
\right),
\label{Hamilton}
\end{equation}
where $m$ denotes the physical mass of the particle. The vectors
$\mathbf{x}=(x_0,x_1,x_2,\dots,x_{P-1})$ and
$\mathbf{v}=(v_0,v_1,v_2,\dots,v_{P-1})$
collect the coordinates and velocities of the beads, respectively.
Cyclic boundary conditions are imposed such that $x_P \equiv x_0$ and $v_P \equiv v_0$.
Here, $V$ denotes the potential energy, and $\omega_P$ is the ring-polymer frequency,
\begin{equation}
\omega_P = \frac{P}{\beta \hbar},
\label{omegaP}
\end{equation}
with $\beta$ being the inverse temperature and $\hbar$ the reduced Planck constant.
In the limit $P \to \infty$, exact quantum-mechanical averages are recovered; at finite
$P$, however, discretization errors arise that depend sensitively on both the system
and the observable considered.

\subsection{Weight functions for harmonic oscillator}

For a one-dimensional harmonic oscillator with mass $m$, the potential energy is
\begin{equation}
V(x) = \frac{1}{2} m \omega_0^2 x^2,
\label{eq:harm_pot}
\end{equation}
where $\omega_0$ is the frequency of the oscillator. The exact quantum-mechanical average total energy is
\begin{equation}
\frac{\hbar\omega_0}{2}\coth\Big(\frac{\beta\hbar\omega_0}{2}\Big). 
\label{eq:E_qm}
\end{equation}

In a PIMD simulation with $P$ beads, the system can be decomposed into normal modes with frequencies
\begin{equation}
\omega_k^2 = \omega_0^2 + 4\omega_P^2 
\sin^2\Big(\frac{\pi k}{P}\Big),
\qquad k = 0, \dots, P-1.
\label{eq:omega_k}
\end{equation}

The classical ring polymer estimator of the internal energy for a given $P$ is
\begin{equation}
    E_P=\frac{1}{\beta}\sum_{k=0}^{P-1} \left( \frac{\omega_0}{\omega_k} \right)^2,
    \label{eq:int_ener}
\end{equation}
which provides the exact expression of Eq.~(\ref{eq:E_qm}) in the $P\rightarrow\infty$ limit. 
The exact result can also be restored for finite $P$ values by reweighting the individual modes, in the frequency domain by $\text{w}_P(\omega_k)$,
such that the sum of the weighted energies matches the exact internal energy of the harmonic oscillator:
\begin{equation}\label{eq:func_eq}
    E_\infty=\frac{1}{\beta}\sum_{k=0}^{P-1} \text{w}_P(\omega_k)\left(\frac{\omega_0}{\omega_k} \right)^2\overset{!}{=}\frac{\hbar\omega_0}{2}\coth\Big(\frac{\beta\hbar\omega_0}{2}\Big).
\end{equation}

The centroid mode can be treated separately, by constraining the frequency-dependent weighting function to $\text{w}_P(\omega_0)=1$:
\begin{equation}\label{eq:func_eq_alt}
    E_\infty=\frac{1}{\beta}+\frac{1}{\beta}\sum_{k=1}^{P-1} \text{w}_P(\omega_k)\left(\frac{\omega_0}{\omega_k} \right)^2\overset{!}{=}\frac{\hbar\omega_0}{2}\coth\Big(\frac{\beta\hbar\omega_0}{2}\Big).
\end{equation}

In the limit $P\to\infty$, $\text{w}_P(\omega_k)\to 1$, ensuring that no correction is applied when the discretization is fully converged. Ceriotti developed a numerical procedure to solve the functional equations of Eqs.~(\ref{eq:func_eq}) and (\ref{eq:func_eq_alt}),\cite{Ceriotti2011-ka} and Brieuc implemented it in CP2K.\cite{Brieuc2016-wy} Recently, an alternative solution was proposed that is more accurate for $P\leq16$, but is not effective for $P>32$.\cite{Madarasz2024-ad} 

In the classical limit ($P = 1$), a number of quantum correction strategies can be understood in terms of a frequency-dependent modification of classical trajectory observables. An early example is the spectral quantum correction approach introduced by Berens and co-workers,\cite{Berens1983-ww} who applied an appropriate quantum weight function to the classical vibrational spectrum in order to obtain quantum-corrected thermodynamic properties from molecular dynamics simulations.

Similarly, the quantum thermal bath (QTB),\cite{Dammak2009-zz} colored-noise thermostat methods,\cite{Ceriotti2009-yq} and GSTA\cite{Berta2020-gz} can all be traced back, in the $P = 1$ limit, to the same harmonic reference weight function defined by Eq.~(\ref{eq:func_eq}). These approaches differ primarily in how the corresponding frequency-dependent modification is realized, either dynamically through thermostat forces acting during the simulation, or \emph{a posteriori} via spectral reweighting or time-domain filtering of the sampled trajectories.

In the context of path-integral-based approaches, this common frequency-dependent target can be realized in different ways depending on how the ring-polymer degrees of freedom are treated. Within this framework, the PI+GLE method and the $f_P^{(0)}$ variant of PIQTB employ a centroid-included formulation Eq.~(\ref{eq:func_eq}), whereas PIGLET and the $f_P^{(1)}$ variant of PIQTB are based on a centroid-excluded formulation Eq.~(\ref{eq:func_eq_alt}). In the present work, we extend GSTA to path-integral simulations using the centroid-included scheme, which allows for a particularly straightforward implementation, as individual ring-polymer replicas can be filtered independently.

\subsection{Extension of GSTA to PIMD simulations}

Having established the common harmonic reference underlying several quantum correction schemes, we now extend this trajectory-based formulation that allows this correction to be applied directly within path-integral simulations.

In GSTA, nuclear quantum effects are incorporated
by applying a frequency-dependent reweighting to trajectory-based observables derived from
molecular dynamics simulations. For harmonic systems, this reweighting can be expressed
analytically and implemented in the time domain via convolution with a suitably defined
kernel function.

We extend this formalism to PIMD by introducing a
bead-number--dependent convolution kernel $g_P(t)$, which accounts for the discretization
error associated with a finite number of path-integral replicas. The kernel is defined as
the inverse Fourier transform of the square root of the frequency-dependent weight
function $\mathrm{w}_P(\omega)$. Since $\sqrt{\mathrm{w}_P(\omega)}$ is an even function of frequency, this
definition can be written explicitly as a cosine transform,
\begin{equation}
g_P(t) = \frac{1}{\pi} \int_{0}^{\infty} \sqrt{\mathrm{w}_P(\omega)} \cos(\omega t)\, d\omega .
\label{eq:kernel}
\end{equation}
Here $\mathrm{w}_P(\omega)$ denotes the same bead-number--dependent weighting function introduced earlier for the harmonic reference system Eqs.~(\ref{eq:func_eq}) and (\ref{eq:func_eq_alt}), now interpreted in the continuous frequency domain.

Given a trajectory-based observable $f(t)$ sampled along a PIMD simulation, such as coordinates, velocities, or forces, its GSTA-filtered counterpart is obtained by applying a bead-number--dependent linear filter in the time domain,
\begin{equation}
\tilde{f}(t) = \left( f \ast g_P \right) (t)
= \int_{-\infty}^{+\infty} f(t - \tau)\, g_P(\tau)\, \mathrm{d}\tau .
\label{eq:convolution}
\end{equation}
The convolution is performed independently for each bead and does not mix different
ring-polymer replicas. In the infinite-bead limit, the kernel $g_P(t)$ reduces to a Dirac
delta function, such that fully converged PIMD trajectories are left unchanged.

In particular, applying Eq.~(\ref{eq:convolution}) to the bead coordinates $x_k(t)$,
velocities $v_k(t)$, and forces $F_k(t)$ yields the filtered quantities
$\tilde{x}_k(t)$, $\tilde{v}_k(t)$, and $\tilde{F}_k(t)$, respectively, which will be used
throughout the following sections.

The GSTA filtering can equivalently be formulated in the frequency domain as a
bead-number--dependent reweighting of the vibrational spectrum. The time-domain convolution
and the frequency-domain reweighting correspond to applying the same frequency-dependent
correction and are related by the convolution theorem. As a consequence, thermodynamic
averages obtained from filtered trajectories are equivalent to those computed from the
reweighted spectrum, provided that consistent estimators are employed.

While the two formulations are equivalent at the level of ensemble averages, the
time-domain representation offers several practical advantages. In particular, it yields
quantum-corrected trajectories defined at each time step (up to short boundary regions at
the beginning and end of the trajectory), enabling time-resolved analysis beyond the scope
of purely spectral approaches. Moreover, the convolution is computationally inexpensive,
as the kernel decays on a timescale of approximately 100~fs at room temperature, making the
method well suited for long trajectories and on-the-fly analysis. 

The formal equivalence between the frequency-domain reweighting and its time-domain convolution representation is shown in Appendix.

\subsection{Energy estimators in standard PIMD and PIGSTA}

We first recall the energy estimators used to define a consistent comparison between
standard PIMD and the PIGSTA framework.

The most straightforward expression for the average kinetic energy $\langle K \rangle$
in PIMD is given by the primitive estimator,
\begin{equation}
 \langle K_{\textnormal{prim}} \rangle
 =
 \frac{m}{2P} \sum_{k=0}^{P-1} \left\langle v_k^2 \right\rangle
 -
 \frac{1}{P} \sum_{k=0}^{P-1}
 \left\langle \frac{1}{2} m \omega_P^2 \left( x_k - x_{k+1} \right)^2 \right\rangle .
\end{equation}
The first term can be evaluated analytically, yielding $1/(2\beta)$.

Within the PIGSTA framework, the same estimator is applied to the filtered trajectories,
\begin{equation}
 \left\langle \tilde{K}_{\textnormal{prim}} \right\rangle
 =
 \frac{m}{2P} \sum_{k=0}^{P-1} \left\langle \tilde{v}_k^2 \right\rangle
 -
 \frac{1}{P} \sum_{k=0}^{P-1}
 \left\langle \frac{1}{2} m \omega_P^2
 \left( \tilde{x}_k - \tilde{x}_{k+1} \right)^2 \right\rangle ,
\end{equation}
where $\tilde{x}_k$ and $\tilde{v}_k$ denote the filtered coordinates and velocities
obtained by convolution with the kernel $g_P(t)$.

The filtering can be applied either on-the-fly during a PIMD simulation or as a
post-processing step. In either case, in the limit $P \to \infty$, the kernel $g_P(t)$ approaches a
Dirac delta function, such that fully converged PIMD trajectories remain unchanged.

Previously it was shown that the application of the filter increases the average kinetic and potential energies by the same amount.\cite{Berta2020-gz} This allows the PIGSTA potential energy to be obtained from an energy-balance relation,
\begin{equation}
 U^{\mathrm{EB}} = \left\langle \tilde{K}_{\textnormal{prim}} \right\rangle - \left\langle K_{\textnormal{prim}} \right\rangle + U_{\textnormal{PIMD}} ,
 \label{eq:UGSTA}
\end{equation}
where $U_{\textnormal{PIMD}}$ denotes the standard PIMD estimator of the potential energy.

Alternatively, the potential energy can be recomputed directly from the filtered
coordinates,
\begin{equation}
 U\left(\tilde{x}\right)
 =
 \sum_{k=0}^{P-1} \frac{V(\tilde{x}_k)}{P} .
 \label{eq:Urec}
\end{equation}
Agreement between $U^{\mathrm{EB}}$ and the ensemble average of $U\left(\tilde{x}\right)$ provides an internal consistency check, indicating that the displacements induced by the filtering remain within the harmonic regime of the potential energy surface.

\subsection{Internally consistent convergence and harmonicity tests}

\subsubsection{Energy test}

The two potential-energy estimators above provide an internally consistent way to assess both the harmonicity of the potential energy surface and the convergence of PIMD with respect to $P$. Large deviations indicate either strong anharmonicity or insufficient bead number.

\subsubsection{Force test}

The same principle can be applied to forces. One may either
(i) convolve the instantaneous forces with the kernel function,
\begin{equation}
\tilde{F}(t) = \left( F \ast g_P \right) (t) ,
\label{eq:F*g}
\end{equation}
or
(ii) compute the forces directly on the filtered coordinates,
\begin{equation}
F(\tilde{x}_i(t)) =
- \frac{\partial V(\tilde{x}_i(t))}{\partial \tilde{x}_i}
- m \omega_P^2 \left( 2\tilde{x}_i(t) - \tilde{x}_{i-1}(t) - \tilde{x}_{i+1}(t) \right) .
\label{eq:Fonfiltx}
\end{equation}
Here, $i$ denotes the bead index (with periodic indexing along the ring polymer), and
$F$ denotes the total ring-polymer force, including both the physical force derived
from the potential $V(x)$ and the harmonic spring forces coupling adjacent beads.

These two force definitions correspond to the two potential-energy estimators introduced above, $U^{\mathrm{EB}}$ and $U(\tilde{x})$, respectively. When the harmonic approximation holds for the displacements induced by the filtering, the forces obtained from the two approaches should coincide. Their relation can be quantified by linear regression and statistical measures such as the coefficient of determination ($R^2$), mean absolute deviation (MAD), or root mean squared deviation (RMSD). Small deviations indicate both harmonicity and convergence with respect to the bead number.

\subsection{Applicability and inherent limitations}

The PIGSTA formalism assumes that the nuclear masses in the PIMD simulation correspond to the physical ones. Although PIMD can be formulated with arbitrary (fictitious) bead masses, such mass scaling changes the normal-mode frequencies and invalidates the harmonic-oscillator-based kernel used in PIGSTA. PIMD codes such as i-PI, CP2K, or LAMMPS typically propagate the ring polymer using the physical nuclear masses.\cite{Litman2024-la,Kuhne2020-zs,Thompson2022-ep}
However, several acceleration schemes—such as staging transformations,\cite{Tuckerman1993-dv} normal-mode and adiabatic PIMD,\cite{Cao1994-ri,Cao1996-np} or adiabatic centroid molecular dynamics (CMD)\cite{Hone2004-wo}— use fictitious masses. Because fictitious masses modify the effective ring-polymer frequencies, the application of PIGSTA in combination with such methods would require corresponding modifications of the weight functions.

The method also relies on the statistical distribution of the ring-polymer normal modes being representative of the target quantum system. Strong thermostat coupling can distort this distribution and lead to an incorrect application of the frequency-dependent weighting. Therefore, thermostats should be chosen so that they minimally perturb the intrinsic statistical properties of the normal modes.

More generally, the PIGSTA framework is formulated for the correction of equilibrium distributions and static observables. Accordingly, the method is not intended to address dynamical properties that depend on real-time correlation functions.

\section{Methods}

\subsection{Zundel cation}

We investigated the H$_5$O$_2^+$ (Zundel) cation at an ultra-low temperature of
1.67~K, closely following the computational protocol of Schran \textit{et al.}\cite{Schran2018-py}
in order to enable a direct comparison of convergence behaviour across different
path-integral acceleration schemes. The simulations employed the high-dimensional
neural network potential developed by Schran \textit{et al.}, which was trained on
CCSD(T*)-F12a/AVTZ reference data\cite{Schran2020-wu} and has been extensively
validated for protonated water clusters.\cite{Schran2018-py,Schran2019-wk}

All PIMD simulations were performed with the CP2K package (version~2023.1).\cite{Kuhne2020-zs}
We considered bead numbers
$P = 2,4,8,16,32,64,128,256,512,1024,$ and $2048$.
For each $P$, the simulations were initiated with a 10~ps equilibration using the
PILE thermostat, with coupling parameters $\tau = 1000$~fs and $\lambda = 0.5$.
Subsequently, a short production run was performed with the thermostat disabled
by setting both $\tau$ and $\lambda$ to zero.

The length of the production run was chosen such that the symmetric convolution
kernel $g_P(t)$ fully spanned the available trajectory, yielding one statistically
independent data point per trajectory.
For $P \leq 1024$, 1000 independent trajectories were generated, while for
$P = 2048$, 60 independent trajectories were performed.
For the reference calculation with $P = 8192$, the same PIMD protocol was used,
with 84 independent trajectories generated due to the computational cost.

The kinetic energy in the PIMD simulations was evaluated using the standard
virial estimator.

\subsection{Liquid water}

Constant-volume (NVT) PIMD simulations were carried out using the MB-pol potential\cite{Babin2013-tp,Babin2014-fd,Medders2014-lw,Reddy2016-pa,Paesani2016-wy}
under periodic boundary conditions.
The simulation cell contained 256 water molecules, corresponding to a cubic box
with a side length of 19.7295~\text{\AA}.

The MB-pol potential was employed with a two-body cutoff of 9.0~\text{\AA},
a three-body cutoff of 4.5~\text{\AA}, a dipole tolerance of $10^{-10}$, and the
conjugate-gradient method for dipole optimization.
All simulations were performed using the LAMMPS software package\cite{Thompson2022-ep} (stable version 2024.09.29), compiled with the MBX~1.2
plugin.\cite{Gupta2025-oq}

All PIMD simulations were propagated with a timestep of 0.25~fs and a local PILE thermostat acting on the ring-polymer beads, with a relaxation
time of $\tau = 100$~fs and a target temperature of $T = 298.15$~K.\cite{Ceriotti2010-zb}
Statistical uncertainties of structural observables were estimated by block averaging, using blocks of 20~ps length.

\subsection{Determination of the weight and kernel functions}

For bead numbers $P \leq 32$, the frequency-dependent weight functions $\mathrm{w}_P(\omega_k)$ were determined using our previously published approach\cite{Madarasz2024-ad}. For $P \geq 64$, the weight functions were obtained using the CP2K implementation for the numerical solution of the functional equation in Eq.~(\ref{eq:func_eq}). The weight functions were determined once for each bead number $P$ in a reduced-frequency representation.

The weight functions were then converted into the corresponding time-domain convolution kernels $g_P(t)$ using the \texttt{w2kernel} program. The kernels were evaluated for the target temperature and the chosen time step. Trajectory post-processing and filtering were performed with the \texttt{wemova} code, which applies the kernels to coordinates, velocities, and forces to yield PIGSTA-corrected observables. For the post-processing of LAMMPS trajectory files, we used our PIGSTA analysis tool in a development version of PLUMED.

\section{Results and Discussion}

We assess the performance of the PIGSTA method in terms of its ability to accelerate the convergence of path-integral molecular dynamics simulations. Before presenting results for specific physical systems, we first examine the properties of the convolution kernels that are central to the PIGSTA approach. This analysis highlights how the kernel shape depends on the number of beads $P$ and on the scheme used for generating the underlying frequency-dependent weights, thereby providing a basis for interpreting the convergence behaviour discussed in the following sections.

The performance of PIGSTA is then evaluated for two representative systems: (i) the Zundel cation (H$_5$O$_2^+$) at ultralow temperature, serving as a prototypical small system with pronounced nuclear quantum effects, and (ii) ambient liquid water as a larger, more complex condensed-phase system. For the Zundel cation, results are compared to conventional PIMD as well as to published benchmarks obtained with established acceleration techniques, namely PIGLET and PIQTB. For liquid water, we focus on a direct comparison between PIGSTA and standard PIMD. The comparison addresses the rate of convergence of kinetic and potential energies, as well as structural observables, as a function of bead number.

In addition, we examine how the internal consistency between formally distinct estimators derived from PIGSTA-filtered trajectories can be used as a reference-free diagnostic of bead-number convergence. This analysis also probes the validity of the underlying harmonic approximation.

\subsection{Kernel functions}
\begin{figure}[H]
\centering
\includegraphics[width=\figwidth]{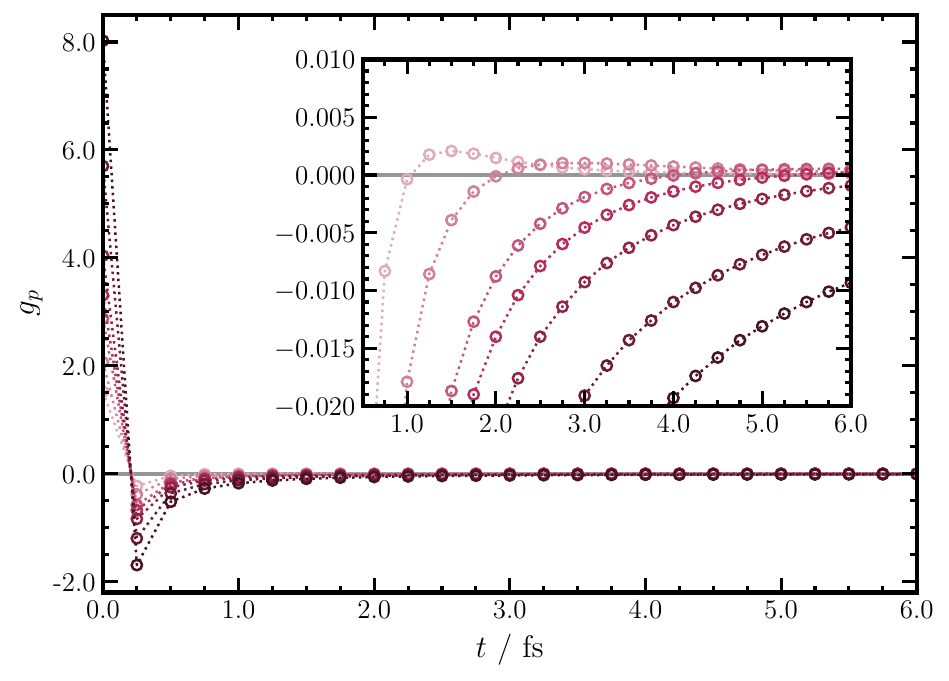}
\caption{\textbf{PIGSTA convolution kernels.}
Convolution kernels $g_P(t)$ evaluated on a discrete time grid for $P=1,2,4,6,8,16,$ and $32$ 
at a temperature of $300$~K using a time step of $0.25$~fs. 
Colors increase in brightness with increasing $P$.}
\label{fgr:kernel_func}
\end{figure}

The convolution kernel $g_P(t)$ encodes the frequency-dependent weighting of PIGSTA in the time domain. Figure~\ref{fgr:kernel_func} illustrates the kernels obtained for different bead numbers $P$. 
The kernel exhibits a singularity at $t=0$, diverging to $+\infty$ exactly at $t=0$ and to $-\infty$ when approaching zero from either side. 
For $P=1$, $g_P(t)$ is negative for all $t \neq 0$ and increases monotonically towards zero. 
For $P>1$, the kernel is negative in the immediate vicinity of the singularity, 
then crosses zero, becomes positive, reaches a finite maximum, and subsequently decays to zero.

The kernel is normalized such that its integral over time equals unity, consistent with its role as a linear convolution filter. 
In practice, the kernels are evaluated on a discrete time grid, such that the divergence at $t=0$ is not explicitly resolved and the kernel values remain finite at the origin. 
With increasing $P$, $g_P(t)$ becomes progressively narrower and approaches a Dirac delta function, indicating that the filtering leaves fully converged PIMD trajectories unchanged in the $P \to \infty$ limit. The decay of the kernel defines the effective temporal window over which trajectory data contribute to the filtered observables. At lower temperatures, the kernels broaden and exhibit a slower temporal decay (not shown).

\subsection{Convergence of thermodynamic properties}

\begin{figure}[H]
\centering
\includegraphics[width=\figwidth]{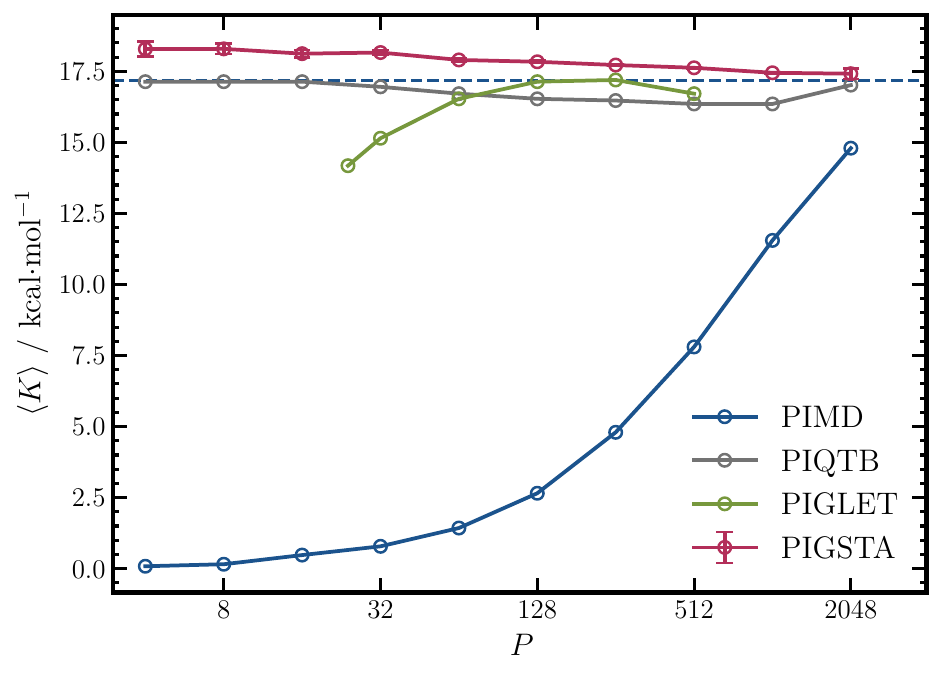}
\caption{\textbf{Kinetic energy convergence of the Zundel cation.}
Convergence of the kinetic energy as a function of the number of beads $P$ at
$T = 1.67$~K. Results from PIGLET and PIQTB simulations are taken from
Schran \textit{et al.}\cite{Schran2018-py}. Error bars indicate 95\% confidence intervals.} 
\label{fgr:kinener}
\end{figure}

We first examine the convergence of the kinetic and potential energies with respect to the number of beads $P$ for the Zundel cation at $T = 1.67$~K. The convergence of the kinetic energy is shown in Figure~\ref{fgr:kinener}.

All accelerated methods (PIGSTA, PIGLET, and PIQTB) rapidly approach a common limiting value for the kinetic energy. In contrast, standard PIMD severely underestimates the kinetic energy over the entire range of bead numbers considered and remains far from convergence even at $P = 2048$. This highlights the dramatic efficiency gains provided by harmonic-reference acceleration schemes for thermodynamic quantities such as the kinetic energy.

\begin{figure}[H]
\centering
\includegraphics[width=\figwidth]{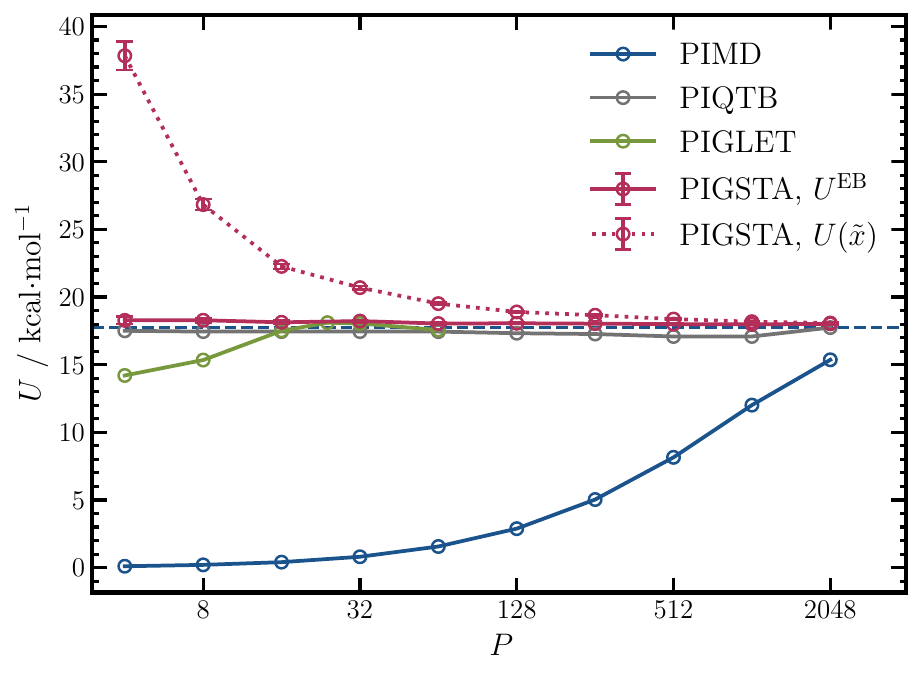}
\caption{\textbf{Potential energy convergence of the Zundel cation.}
Convergence of the potential energy as a function of the number of beads $P$. For PIGSTA, both the energy-balance estimator $U^{\mathrm{EB}}$ and the recomputed potential energy on the filtered coordinates $U(\tilde{x})$ are shown. PIQTB and PIGLET data are taken from Schran \textit{et al.}\cite{Schran2018-py}. Error bars indicate 95\% confidence intervals.} 
\label{fgr:potener}
\end{figure}

The convergence of the potential energy is shown in Figure~\ref{fgr:potener}. For standard PIMD, PIGLET, and PIQTB, the potential energy approaches a common limiting value as the number of beads is increased. For PIGSTA, we consider two estimators: the energy-balance estimator $U^{\mathrm{EB}}$, obtained from the kinetic energy balance Eq.~(\ref{eq:UGSTA}), and the recomputed potential energy on the filtered coordinates, $U(\tilde{x})$ Eq.~(\ref{eq:Urec}). At sufficiently large $P$, both PIGSTA estimators converge to the same limiting value as the other methods within statistical uncertainty.

At low bead numbers, marked differences emerge between the two PIGSTA estimators. While the energy-balance estimator $U^{\mathrm{EB}}$ already provides a reasonable estimate of the potential energy, substantially improving upon the strong underestimation observed in standard PIMD, the recomputed estimator $U(\tilde{x})$ converges much more slowly. This indicates that $U^{\mathrm{EB}}$ provides the more robust finite-$P$ estimate, whereas deviations in $U(\tilde{x})$ offer a sensitive internal diagnostic of both anharmonicity and an insufficient number of beads. The agreement of the two estimators at larger bead numbers therefore indicates that the filtered displacements remain within the locally harmonic regime of the potential energy surface. We note that established acceleration techniques such as PIGLET and PIQTB similarly rely on harmonic reference assumptions and are formally exact only in the harmonic limit.

\subsection{Convergence of structural properties}

We next examine the convergence of structural observables of the Zundel cation, focusing on quantities that are well known to be highly sensitive to nuclear quantum effects. Specifically, we consider (i) the proton-sharing coordinate $\delta$ and (ii) the gyration radius of the shared proton, $r_{\mathrm{gyr}}$. Both quantities are evaluated directly from the PIGSTA-filtered coordinates.

The proton-sharing coordinate is defined as\begin{equation}
\delta = r_{\textnormal{OH}_1} - r_{\textnormal{OH}_2},
\label{eq:delta_def}
\end{equation}
where $r_{\textnormal{OH}_1}$ and $r_{\textnormal{OH}_2}$ denote the distances between the shared proton and the two oxygen atoms. This coordinate measures the instantaneous asymmetry of the proton position between the two potential wells.

\begin{figure}[H]
\centering
\includegraphics[width=\figwidth]{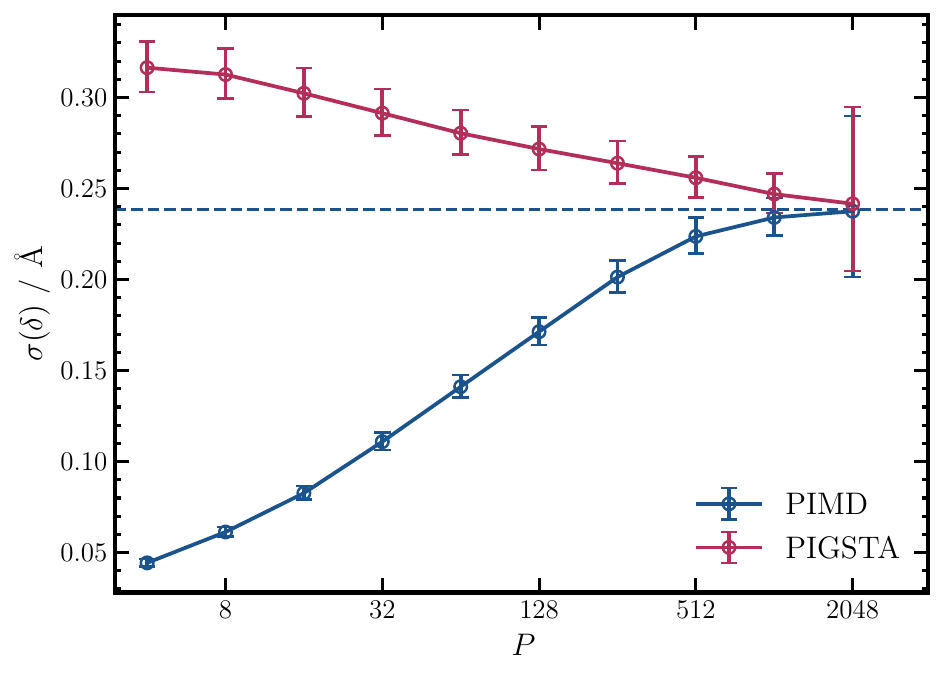}
\caption{\textbf{Convergence of the proton-sharing coordinate fluctuations.}
Convergence of the standard deviation of the proton-sharing coordinate,
$\sigma(\delta)$, as a function of the number of beads $P$.
Only PIMD and PIGSTA values are shown. PIGSTA values are obtained directly from the
filtered coordinates. Error bars indicate 95\% confidence intervals.}
\label{fgr:delta}
\end{figure}

Figure~\ref{fgr:delta} shows the convergence of the standard deviation of the proton-sharing coordinate, $\sigma(\delta)$, as a function of bead number $P$. At low $P$, standard PIMD severely underestimates $\sigma(\delta)$, whereas PIGSTA yields substantially improved estimates up to intermediate bead numbers ($P \leq 256$), bringing the fluctuations much closer to the converged quantum limit. For $P \geq 512$, the remaining deviations of PIMD and PIGSTA become comparable in magnitude, with the two approaches converging from opposite directions. Full agreement is only reached at the largest bead number considered ($P = 2048$), underscoring the particularly slow convergence of this structural measure of proton sharing.

The gyration radius of the shared proton provides a complementary measure of nuclear quantum delocalisation. It is defined as
\begin{equation}
r_{\mathrm{gyr}}^2 
  = \frac{1}{P} \sum_{k=0}^{P-1}
    \left| \mathbf{r}_k - \mathbf{r}_{\mathrm{c}} \right|^2,
\qquad
\mathbf{r}_{\mathrm{c}} = \frac{1}{P} \sum_{k=0}^{P-1} \mathbf{r}_k,
\label{eq:gyr_def}
\end{equation}
where $\mathbf{r}_k$ denote the filtered bead coordinates of the proton and
$\mathbf{r}_{\mathrm{c}}$ their centroid.

\begin{figure}[H]
\centering
\includegraphics[width=\figwidth]{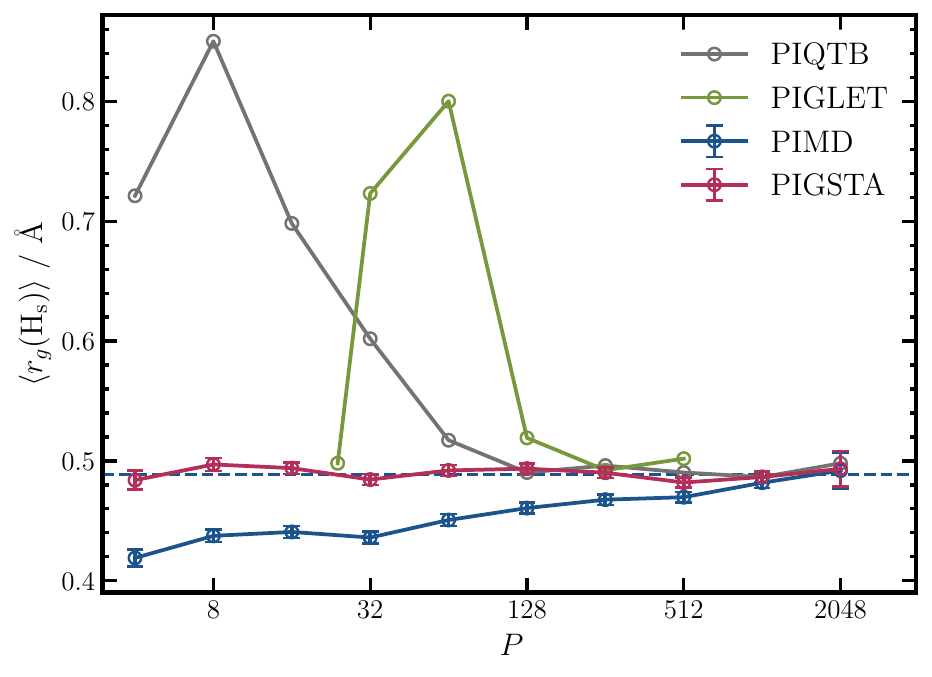}
\caption{\textbf{Gyration radius of the shared proton.}
Convergence of the gyration radius $r_{\mathrm{gyr}}$ as a function of the number of beads $P$. PIGSTA values are computed from filtered coordinates and compared to PIMD, PIGLET, and PIQTB. Reference PIGLET and PIQTB data are taken from Schran \textit{et al.}\cite{Schran2018-py}. Error bars indicate 95\% confidence intervals.}
\label{fgr:gyrHs}
\end{figure}

Figure~\ref{fgr:gyrHs} shows the convergence of the gyration radius $r_{\mathrm{gyr}}$ as a function of the number of beads $P$ for PIMD, PIGLET, PIQTB, and PIGSTA. All methods converge to the same limiting value at sufficiently large $P$. At low bead numbers, standard PIMD significantly underestimates $r_{\mathrm{gyr}}$, indicating an artificially localized description of the shared proton. In contrast, PIGSTA yields values much closer to the converged quantum limit already at small $P$, substantially reducing the finite-bead discretization error for this measure of spatial delocalisation. The PIGLET and PIQTB results exhibit pronounced non-monotonic behaviour at intermediate bead numbers, with large overshoots relative to the converged value, before relaxing at larger $P$.

Overall, these trends are consistent with those observed for the proton-sharing coordinate and further highlight the particularly slow bead-number convergence of structural measures of proton delocalisation compared to thermodynamic quantities. Taken together, they demonstrate that PIGSTA substantially reduces finite-$P$ discretization errors across the observables considered, thereby accelerating bead-number convergence relative to standard PIMD.

\subsection{Force-based convergence and harmonicity test}

The energy-based analysis in the previous sections provides a first indication of convergence, but it may partially mask error compensation between different contributions to the potential energy surface. To obtain a more sensitive and local diagnostic, we therefore analyse the forces as well. Two force estimators are considered: the kernel-convolved forces $\widetilde{F}(x(t))$ Eq.~(\ref{eq:F*g}) and the forces evaluated on the filtered coordinates $F(\widetilde{x}(t))$ Eq.~(\ref{eq:Fonfiltx}). In the harmonic limit, these two force estimators are expected to coincide, such that their relationship can be quantified by a linear fit. Here, $F$ denotes the total ring-polymer force, including both physical and spring contributions.

\begin{figure}[H]
\centering
\includegraphics[width=\figwidth]{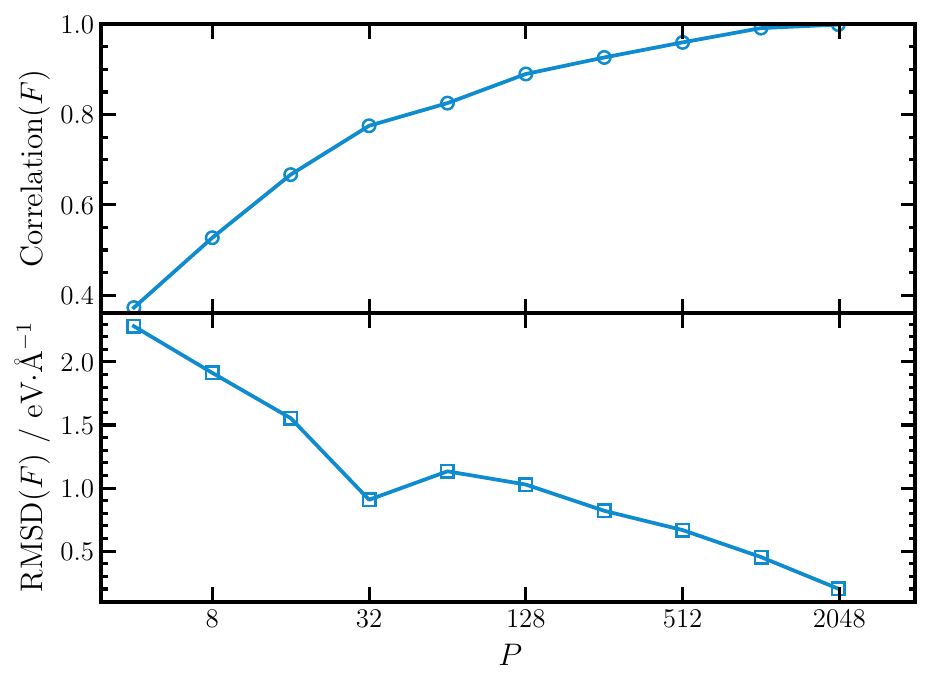}
\caption{\textbf{Force-based convergence and harmonicity diagnostics.}
Quality of the force-based PIGSTA estimators for the Zundel cation as a function of the number
of beads $P$. Top: coefficient of determination $R^2$ from linear fits between the
kernel-convolved forces $\widetilde{F}(x(t))$ and the forces evaluated on the
filtered coordinates $F(\widetilde{x}(t))$. Bottom: root mean squared deviation
(RMSD) between the two force estimators.}
\label{fgr:for_dev}
\end{figure}

Figure~\ref{fgr:for_dev} summarises this comparison for the Zundel cation at $T = 1.67$~K. The upper panel shows the coefficient of determination $R^2$ obtained from linear fits between $\widetilde{F}(x(t))$ and $F(\widetilde{x}(t))$ as a function of the number of beads $P$. The correlation increases systematically with increasing $P$ and approaches unity for $P \ge 1024$, indicating that the two force estimators become strongly correlated. The corresponding slope of the fit (not shown) approaches one in the same regime, consistent with an increasingly harmonic relation between the filtered coordinates and the underlying forces. At very large bead numbers, however, the ring-polymer spring forces contribute similarly to both estimators, such that correlation alone becomes less informative.

To obtain a more robust measure, the lower panel of Figure~\ref{fgr:for_dev} reports the root mean squared deviation (RMSD) between $\widetilde{F}(x(t))$ and $F(\widetilde{x}(t))$. The RMSD decreases overall with increasing $P$ and reaches its minimum at $P = 2048$, which we take as the most reliable PIGSTA reference in the present study. A non-monotonic behaviour is observed at intermediate bead numbers, with a smaller RMSD at $P = 32$ than at $P = 128$. This can be traced back to the different construction of the underlying weight functions: for $P = 32$ they were obtained using the deconvolution-based procedure,\cite{Madarasz2024-ad} whereas for $P \ge 64$ the weights were taken directly from CP2K PIQTB output.

Taken together, the force-based analysis confirms that (i) PIGSTA results obtained at intermediate-to-large bead numbers, and in particular at $P = 2048$, are the most reliable.
Moreover, the agreement between the two force estimators provides an internally consistent, reference-free diagnostic of both bead-number convergence and the validity of the harmonic approximation, complementing the energy-based tests.

\subsection{Structural convergence in liquid water}

\begin{figure}[H]
    \centering
    \includegraphics[width=0.9\linewidth]{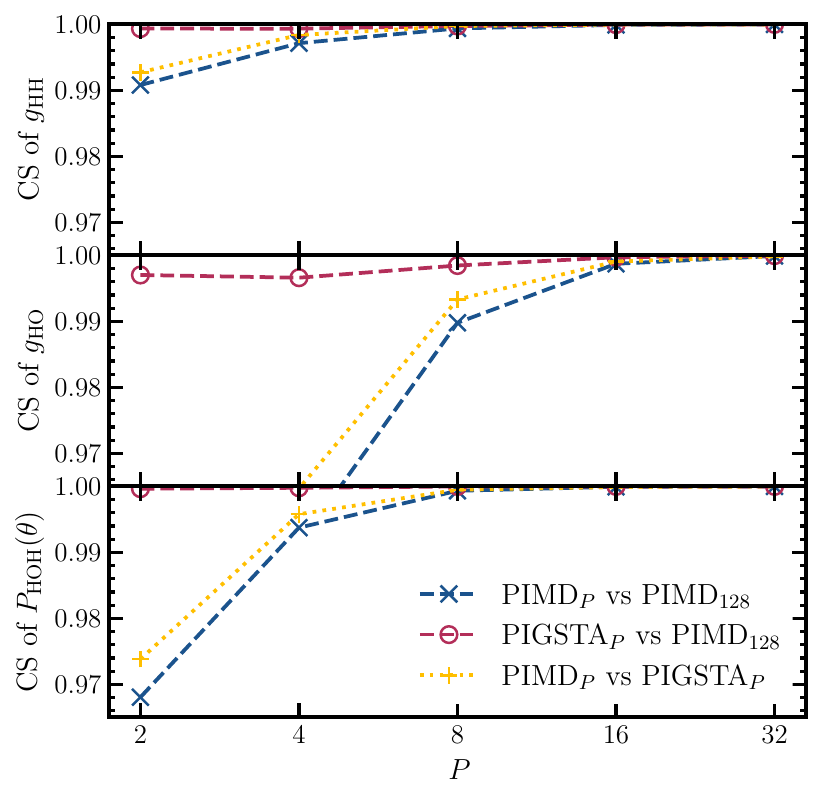}
    \caption{\textbf{Structural convergence of liquid water.} Cosine similarity (CS) between RDFs and ADFs obtained at bead number $P$ and a reference PIMD simulation with $P = 128$ for H--H, H--O pair-correlations, and H--O--H angles. Results are shown for both PIMD and PIGSTA, together with their relative similarity at each $P$.}
    \label{fig:gradfscores}
\end{figure}

To assess the performance of PIGSTA beyond the ultralow-temperature Zundel cation,
we also examine its behaviour for ambient liquid water, a widely used benchmark system
in which several structural observables exhibit only gradual convergence with respect
to the number of beads.\cite{Ceriotti2016-no,Yao2021-bl,Daru2022-rj,Stolte2024-tj,Meszaros2025-xf} Figure~\ref{fig:gradfscores} quantifies the consistency between radial distribution functions (RDFs) and angular distribution functions (ADFs) obtained at various bead numbers and a reference PIMD curve at $P = 128$, using the cosine similarity (CS) metric. Results are shown for the quantities most sensitive to nuclear quantum effects: namely H--H, H--O pair correlations, and H--O--H angles.

The H--H RDF, which exhibits moderate nuclear quantum effects, converges rapidly
with respect to $P$ for both PIMD and PIGSTA, with PIGSTA closely matching the reference already at very small bead numbers. For the H--O pair, where nuclear quantum effects are more pronounced, convergence is noticeably slower. 

\begin{figure}[H]
    \centering
    \includegraphics[width=0.9\linewidth]{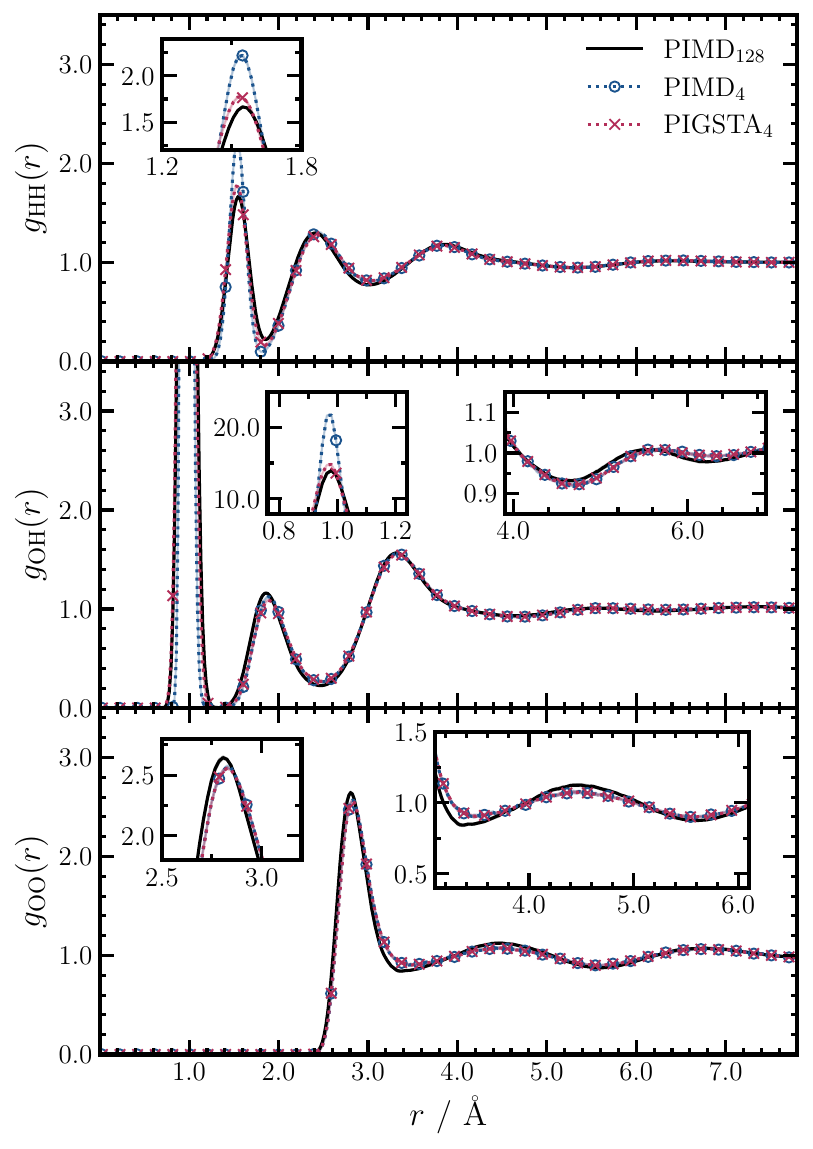}
    \caption{\textbf{Radial distribution functions at small bead number.} H--H, H--O, and O--O RDFs of liquid water obtained with PIMD and PIGSTA at $P = 4$, compared to the converged PIMD reference at $P = 128$. Shaded areas indicate the standard error of the mean estimated by block averaging using 20 ps long blocks, and a bin width of 0.02 \r{A}.}
    \label{fig:grs}
\end{figure}

In this case, PIGSTA systematically yields higher similarity with the reference than standard PIMD at small $P$, and both methods approach near-complete overlap by $P = 32$ within the resolution of the CS metric.  Additional insight into the local structure of liquid water is provided by angular distribution functions (ADFs). Among these, the H--O--H angle is the most sensitive to proton delocalisation. In this case, PIGSTA reaches near-complete agreement with the reference distribution already at small bead numbers ($P \approx 2$--4, within the resolution of the CS metric). In contrast, standard PIMD requires at least 8 beads to achieve a similar level of consistency. 

\begin{figure}[H]
    \centering
    \includegraphics[width=0.9\linewidth]{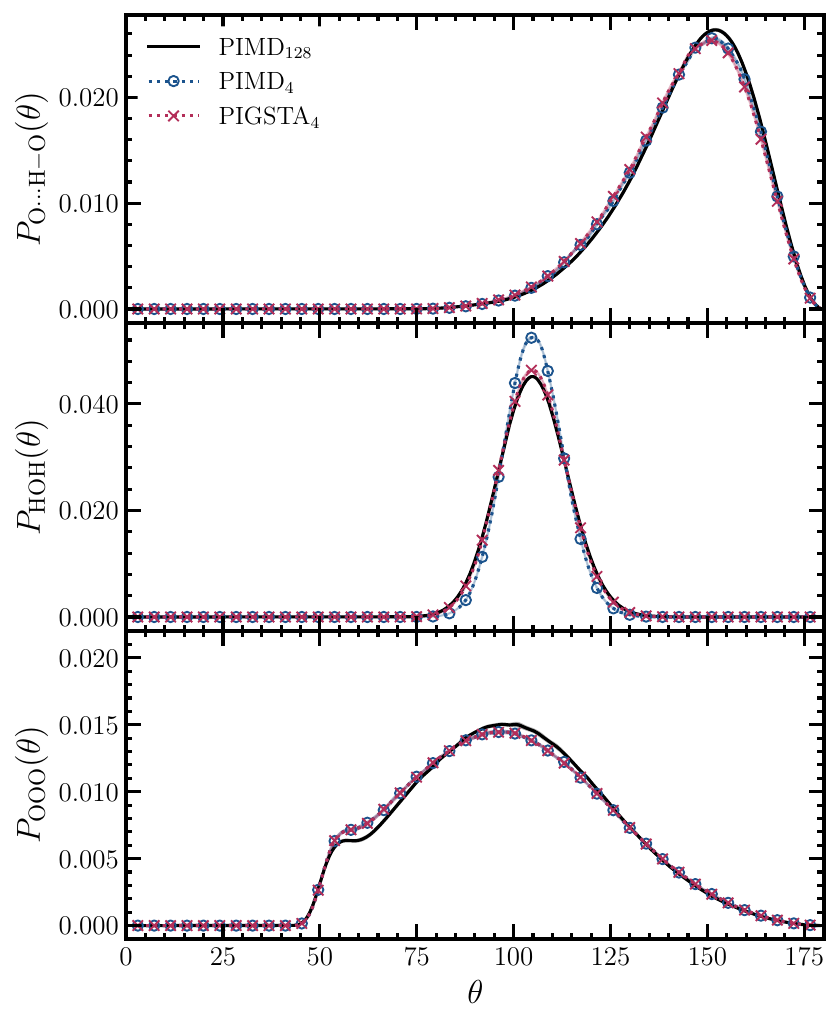}
    \caption{\textbf{Angular distribution functions at small bead number.} O$\cdots$H--O, H--O--H, and O--O--O angular distributions in degrees of liquid water obtained with PIMD and PIGSTA at $P = 4$, compared to the converged PIMD reference at $P = 128$. Shaded areas indicate the standard error of the mean estimated by block averaging using 20 ps long blocks, and a bin width of 2.0$^{\circ}$.}
    \label{fig:adfs}
\end{figure}

Figures~\ref{fig:grs} and \ref{fig:adfs} provide a direct comparison of the RDFs and ADFs obtained with PIMD and PIGSTA at $P = 4$. The visual agreement with the converged reference is consistent with the trends observed in the cosine similarity analysis. For the O--O, O--O--O and O$\cdots$H--O correlations, nuclear quantum effects are hardly noticeable; PIMD and PIGSTA display nearly identical behaviour, showing that PIGSTA does not introduce artificial distortions of the intermolecular structure. Overall, these results indicate that PIGSTA accelerates the convergence of structural correlation functions in liquid water at small bead numbers, while remaining fully consistent with standard PIMD in the converged regime.

\section{Conclusions}

In this work, we introduced PIGSTA, a post-processing extension of generalized smoothed
trajectory analysis to path-integral molecular dynamics. By filtering path-integral
trajectories with analytically defined convolution kernels derived from the quantum
harmonic-oscillator model, PIGSTA incorporates nuclear quantum fluctuations without
modifying the underlying dynamics and remains exact in the harmonic limit. As a result,
quantum fluctuations can be recovered consistently even from discretized path-integral
representations.

From a practical perspective, PIGSTA directly addresses the bead-number convergence
problem in PIMD by improving the convergence of both thermodynamic and structural
observables at reduced numbers of replicas. As a parameter-free, post-processing
approach, it alleviates a major computational bottleneck without requiring
system-specific tuning, prior knowledge of the vibrational spectrum, or any modification
of the simulation protocol.

Within its intended scope of application, PIGSTA can be safely applied to
path-integral molecular dynamics simulations and should be regarded as a generally
beneficial post-processing step. Beyond improving numerical convergence, a central
feature of PIGSTA is that it enables the assessment of convergence in a physically
meaningful sense. By comparing formally distinct estimators obtained from filtered
trajectories, the method provides an internal, reference-free diagnostic of whether the
sampled configurations remain consistent with the local shape of the underlying
potential energy surface. When this condition is satisfied, the resulting quantum
fluctuations can be interpreted physically, and different path-integral acceleration
schemes are expected to yield mutually consistent results. This highlights the value of
internal consistency as a practical and reliable criterion for assessing the physical
validity of path-integral simulations.

This diagnostic capability is particularly important in extreme regimes, such as the
ultralow-temperature Zundel cation, where bead-number convergence can be strongly
non-monotonic and property-dependent. In this regime, PIGSTA provides a transparent
framework to identify when reduced bead representations remain reliable and when
increasing the number of beads is necessary to restore physical consistency.

By contrast, for ambient liquid water at room temperature, bead-number convergence is
achieved at substantially smaller numbers of replicas. In this regime, PIGSTA yields
structural observables that are fully consistent with standard PIMD already at moderate
bead numbers, demonstrating that the method does not introduce artifacts and remains
compatible with established benchmark descriptions of liquid water.

Taken together, these results demonstrate that PIGSTA provides a practical and reliable
extension of path-integral molecular dynamics for the treatment of nuclear quantum
effects in complex systems. As a post-processing framework applicable to both classical
and path-integral molecular dynamics trajectories, PIGSTA combines accelerated
convergence with an internal assessment of the physical consistency of quantum
fluctuations. This dual capability enables efficient simulations across a wide range of
temperatures and system sizes, while providing a transparent criterion for determining
when reduced bead representations remain physically meaningful. Accordingly, for any
PIMD simulation that falls within the methodological framework of PIGSTA, applying this
post-processing step represents a no-regret strategy: converged results remain
unchanged, while unconverged simulations either benefit from improved estimates or are
identified as internally inconsistent at the chosen bead number.

\section{Appendix}
\subsection{Formal basis of PIGSTA}
\label{app:PIGSTA}

This Appendix establishes the formal equivalence between the frequency-domain reweighting underlying PIGSTA and its time-domain convolution formulation, which is used throughout the main text.

\subsubsection{Harmonic reference and trajectory-based spectra}
\label{app:harmonic_reference}

We consider a single harmonic oscillator as a reference system within the harmonic
approximation, defined by the Hamiltonian given in Eq.~(\ref{eq:harm_pot}) of the main text. This model captures the essential features of the path-integral discretization and provides a well-defined setting for analyzing trajectory-based spectral quantities.

Starting from a classical MD or PIMD simulation, the coordinate power spectral density $S_{xx}(\omega)$ is defined as
\begin{equation}
S_{xx}(\omega)=
\lim_{\tau\to\infty}\frac{1}{2\tau}
\left|
\mathcal{F}_\tau\{x\}(\omega)
\right|^2 ,
\label{eq:Sxx}
\end{equation}
where $\mathcal{F}_\tau\{x\}(\omega)$ denotes the Fourier transform of the trajectory
over a finite time interval $[-\tau,\tau]$. Explicitly, the Fourier transform of a time-dependent quantity $f(t)$ is defined as
\begin{equation}
\mathcal{F}\{f\}(\omega)=\int_{-\infty}^{\infty}
f(t)\,\mathrm{e}^{-i\omega t}\,\mathrm{d}t .
\end{equation}
For notational simplicity, we omit the explicit $\tau$ dependence of the finite-time Fourier transforms in the following.

For classical MD, the resulting power spectral density $S_{xx}(\omega)$ provides a direct representation of the vibrational content of the system within the harmonic approximation. In the context of PIMD, however, the sampled trajectories correspond to an extended ring-polymer representation with $P$ beads, and the associated $S_{xx}(\omega)$ contains contributions from both physical vibrational modes and internal ring-polymer modes.

As a consequence, the power spectral density obtained directly from finite-bead PIMD
trajectories does not admit a direct physical interpretation in terms of the true
vibrational frequencies of the underlying quantum system, but should instead be
regarded as a well-defined spectral representation of the corresponding harmonic
reference system at finite path-integral discretization.

For a harmonic oscillator, the spectrum consists of discrete peaks located at the normal-mode frequencies $\omega_k$ of the ring-polymer representation. The amplitudes of these peaks scale as $\propto \omega_k^{-2}$, up to convention-dependent prefactors, in accordance with Eqs.~(\ref{eq:omega_k}) and (\ref{eq:int_ener}) of the main text. Within this framework, the internal energy of a harmonic oscillator at finite bead number $P$ can be expressed in terms of the coordinate power spectral density using Parseval’s theorem as
\begin{equation}
E_P
= m \omega_0^2 \langle x^2 \rangle
= \frac{1}{\pi} m \omega_0^2
\int_0^{\infty}
S_{xx}(\omega)\,
\mathrm{d}\omega .
\end{equation}

\subsubsection{Frequency-domain reweighting of the PIMD spectrum}
\label{app:frequency_reweighting}

Within the harmonic approximation, the exact quantum-mechanical internal energy can be
recovered from a finite-bead PIMD simulation by applying a frequency-dependent reweighting to the corresponding coordinate power spectral density. Formally, this corresponds to multiplying $S_{xx}(\omega)$ by a bead-number--dependent weight function
$\mathrm{w}_P(\omega)$ and integrating over all frequencies,
\begin{equation}
E_\infty
=
\frac{1}{\pi} m \omega_0^2
\int_0^{\infty}
\mathrm{w}_P(\omega)\,
S_{xx}(\omega)\,
\mathrm{d}\omega .
\label{eq:ESxx}
\end{equation}

The weight function $\mathrm{w}_P(\omega)$ coincides with the function defined in
Eq.~(\ref{eq:func_eq}) of the main text and provides the bead-number--dependent correction that maps the finite-bead classical spectrum onto the exact quantum-mechanical internal energy $E_\infty$ of the harmonic reference system.

\subsubsection{Time-domain formulation and convolution kernel}
\label{app:convolution_kernel}

Direct reweighting in the frequency domain is inconvenient for practical applications, as it requires explicit computation and manipulation of spectral quantities. The frequency-dependent reweighting can instead be recast in the time domain by exploiting the equivalence between multiplication in the frequency domain and convolution in the time domain.

Specifically, multiplying the coordinate power spectral density by the weight function $\mathrm{w}_P(\omega)$ is equivalent to applying a linear filter to the corresponding time-domain trajectory. For coordinates, the filtered trajectory can be written as
\begin{equation}
\tilde{x}(t)
=
\int_{-\infty}^{\infty}
g_P(t - \tau)\, x(\tau)\, \mathrm{d}\tau = (g_P*x)(t),
\label{eq:coordinate_convolution}
\end{equation}
where $g_P(t)$ is a bead-number--dependent convolution kernel. While coordinates are used here for definiteness, the same construction applies to any trajectory-based vibrational observable.

The kernel $g_P(t)$ is defined as the inverse Fourier transform of the square  root of the
weight function,
\begin{equation}
g_P(t)
=
\frac{1}{2\pi}
\int_{-\infty}^{\infty}
\sqrt{\mathrm{w}_P(\omega)}\,
\mathrm{e}^{i \omega t}\,
\mathrm{d}\omega .
\label{eq:kernel_definition}
\end{equation}
Therefore, 
\begin{equation}
\mathrm{w}_P(\omega)=  \left|\mathcal{F}\{g_P\}(\omega)\right|^2 . 
\label{eq:wP}
\end{equation}
Since $\mathrm{w}_P(\omega)$ is a real and even function of frequency, the above definition is equivalent to the cosine-transform representation given in the main text Eq.~(\ref{eq:kernel}).

Plugging Eq. (\ref{eq:Sxx}) and Eq. (\ref{eq:wP}) into Eq. (\ref{eq:ESxx}) we obtain
\begin{align}
E_\infty&=
\frac{1}{\pi} m \omega_0^2
\int_0^{\infty}
\left|\mathcal{F}\{g_P\}(\omega)\right|^2 \left|\mathcal{F}\{x\}(\omega)\right|^2 
\mathrm{d}\omega\\ 
&=
\frac{1}{\pi} m \omega_0^2
\int_0^{\infty}
\left|\mathcal{F}\{g_P * x\}(\omega)\right|^2 
\mathrm{d}\omega\\ 
&=\frac{1}{\pi} m \omega_0^2
\int_0^{\infty}
\left|\mathcal{F}\{\tilde{x}\}(\omega)\right|^2 
\mathrm{d}\omega\\
&= \frac{1}{\pi} m \omega_0^2
\int_0^{\infty}
S_{\tilde{x}\tilde{x}}(\omega)\,
\mathrm{d}\omega\\
&=m \omega_0^2 \langle \tilde{x}^2 \rangle,
\end{align}
where we used the convolution theorem in the second equality and $S_{\tilde{x}\tilde{x}}(\omega)$ is defined analogously to Eq.~(\ref{eq:Sxx}) for the filtered trajectory $\tilde{x}(t)$.

In practice, when implemented numerically, $g_P(t)$ inherits the temperature dependence of the weight function $\mathrm{w}_P(\omega)$ and, when evaluated numerically, also depends on the chosen time discretization through the finite integration step. This time-domain formulation makes it possible to apply the bead-number--dependent correction entirely in post-processing, without modifying the underlying equations of motion, thermostatting scheme, or sampling procedure.

As a direct consequence of the differentiation rules of convolution, the same filtering procedure can be consistently applied not only to the coordinates, but also to the corresponding velocities and forces. In particular, time derivatives commute with the convolution operator defined by $g_P(t)$, such that filtering the coordinates and then differentiating is equivalent to filtering the corresponding time derivatives. Together, these properties enable a unified treatment of trajectory-based observables within the present time-domain formulation and provide the formal basis for the energy- and force-consistency tests introduced in the main text, enabling a reference-free assessment of bead-number convergence and local harmonicity.

\section*{Declarations}

\begin{itemize}
\item \textbf{Data availability}
The datasets supporting the findings of this work, including simulation outputs for the Zundel cation and processed RDF/ADF observables for ambient liquid water, are available on Zenodo at \url{https://doi.org/10.5281/zenodo.18302429}. Precomputed weight functions used in this work are also provided via Zenodo for $P = 2, 4, 6, 8, 16,$ and $32$ (\url{https://doi.org/10.5281/zenodo.10702413}), and for $P = 64, 128, 256, 512, 1024,$ and $2048$ (\url{https://doi.org/10.5281/zenodo.18255071}).
\item \textbf{Code availability}
Input files and analysis scripts used for the PIMD simulations and the PIGSTA post-processing of the Zundel cation are available at \url{https://github.com/madaraszadam/Zundel_PIMD_PIGSTA}. 
The w2kernel and wemova tools used for kernel generation and trajectory filtering are available at \url{https://github.com/madaraszadam/w2kernel} and \url{https://github.com/madaraszadam/wemova}, respectively. A development implementation of (PI)GSTA within PLUMED is available at \url{https://github.com/madaraszadam/PLUMED-GSTA}.
\item \textbf{Acknowledgements}
We thank Professor Dominik Marx and Professor Christoph Schran for granting access to the ML-PES of the Zundel cation. This paper was supported by the János Bolyai Research Scholarship of the Hungarian Academy of Sciences, by the National Research, Development and Innovation Office of Hungary (NKFIH, Grant No.  FK142784 and FK147031, DKOP-23), and the ELTE University Excellence Fund. We acknowledge Digital Government Development and Project Management Ltd. for awarding us access to the Komondor HPC facility based in Hungary.
\item \textbf{Author contributions} 
JD motivated the extension of GSTA to PIMD, and identified suitable benchmark systems. AM derived and implemented the PIGSTA methodology, led the project and performed the simulations for the Zundel cation. BM performed the liquid water simulations under the supervision of JD. All authors analyzed and interpreted the data. BM prepared the figures. All authors contributed to the writing of the manuscript and approved the final version.
\item \textbf{Competing interests} The authors declare no competing financial or non-financial interests. 
\end{itemize}

\printbibliography
\end{multicols}
\end{document}